\documentclass{aa}
\usepackage{graphicx}
\usepackage{longtable}
\usepackage[varg]{txfonts}
\begin{document}
   \title{Optical photometric and spectral study of the new FU Orionis object V2493 Cyg (HBC 722)\thanks{Tables 1 and 3 are only available in electronic form at the CDS via anonymous ftp to cdsarc.u-strasbg.fr (130.79.128.5) or via http://cdsweb.u-strasbg.fr/cgi-bin/qcat?J/A+A/}}


   \author{E. H. Semkov\inst{1}
          \and    
          S. P. Peneva\inst{1}
          \and
          U. Munari\inst{2}
          \and
          M. K. Tsvetkov\inst{1}
          \and
          R. Jurdana-\v{S}epi\'{c}\inst{3}
          \and
          E. de Miguel\inst{4}
          \and
          R. D. Schwartz\inst{5}
          \and
          D. P. Dimitrov\inst{1}
          \and
          D. P. Kjurkchieva\inst{6}
          \and
          V. S. Radeva\inst{6,7}}

   \offprints{E. Semkov (esemkov@astro.bas.bg)}

   \institute{Institute of Astronomy and National Astronomical Observatory, Bulgarian Academy of Sciences,
              72 Tsarigradsko Shose blvd., BG-1784 Sofia, Bulgaria,
              \email{esemkov@astro.bas.bg}
         \and
         INAF Osservatorio Astronomico di Padova, Sede di Asiago, I-36032 Asiago (VI), Italy
         \and
         Physics Department, University of Rijeka, Omladinska 14, HR-51000 Rijeka, Croatia
         \and
         Departamento de Fisica Aplicada, Facultad de Ciencias Experimentales, Universidad de Huelva, 21071 Huelva, Spain
         \and
         Galaxy View Observatory, 102 Galaxy View Ct., Sequim, WA 98382, USA
         \and
         Department of Physics, Shumen University, 9700 Shumen, Bulgaria
         \and
         Public Astronomical Observatory and Planetarium, Varna, Bulgaria}

   \date{Received ; accepted }


  \abstract
   {}
   {We present new results from optical photometric and spectroscopic observations of the eruptive pre-main sequence star V2493 Cyg (HBC 722). 
   The object has continued to undergo significant brightness variations over the past few months and is an ideal target for follow-up observations.}
   {We carried out CCD $BVRI$ photometric observations in the field of V2493 Cyg ("Gulf of Mexico") from August 1994 to April 2012, i.e. at the pre-outburst states and during the phases of the outburst. 
   We acquired high, medium, and low resolution spectroscopy of V2493 Cyg during the outburst. 
   To study the pre-outburst variability of the target and construct its historical light curve, we searched for archival observations in photographic plate collections. 
   Both CCD and photographic observations were analyzed using 15 comparison stars in the field of V2493 Cyg.}
   {The pre-outburst photographic and CCD photometric observations of V2493 Cyg show low-amplitude light variations typical of T Tauri stars. 
   The recent photometric data show a slow light decrease from October 2010 to June 2011 followed by an increase in brightness that continued until early 2012. 
   The spectral observations of V2493 Cyg are typical of FU Orionis stars absorption spectra with strong P Cyg profiles of H$\alpha$ and Na I D lines.
   On the basis of photometric monitoring performed over the past two years, the spectral properties at the maximal light, as well as the shape of long-term light curves, we confirm that the observed outburst of V2493 Cyg is of FU Orionis type.}
   {}


   \titlerunning{Optical photometric and spectral study of V2493 Cyg}
   \maketitle
%

\section{Introduction}

Studies of the photometric and spectroscopic variability of pre-main sequence (PMS) stars are very important to improving our understanding of the early stages of stellar evolution.
One of the most remarkable PMS phenomena associated with a significant increase in brightness is the FU Orionis (FUor) outburst. 
This corresponds to young eruptive stars as originally classified by Herbig (1977) following the discovery of the FUor objects V1057 Cyg and V1515 Cyg.
Several more objects have since been assigned to this class of young variables (see Reipurth \& Aspin 2010 and references therein).
All known FUors share the same defining characteristics: a $\Delta$$V$$\approx$4-6 mag outburst amplitude, association with reflection nebulae, location in star-forming regions, an F-G supergiant spectrum during outbursts, a strong LiI~6707~\AA\ line in absorption, and CO bands in near-infrared spectra (Herbig 1977, Reipurth \& Aspin 2010).  
An important feature of FUors is the massive supersonic wind observed as a P Cyg profile most commonly for both H$\alpha$ and Na I D lines.
A typical outburst of FUor objects can last for several decades, and the rise time is shorter than that of the decline.

Herbig (1989) also defined another class of young eruptive stars characterized by more than one outburst in the optical light, and referred to them as EXors after the prototype of this class, the PMS star EX Lupi. 
These EXor objects undergo frequent (every few years or decades), irregular, and relatively brief (a few weeks to a few months or one year) outbursts with amplitudes $\Delta$$V$$\approx$2-5. 
During these events, the cool spectrum of the quiescence is veiled and strong emission lines from single ionized metals appear with reversed P-Cyg absorption components (Herbig 2007; Aspin et al. 2010). 
Both types of eruptive stars, FUors and EXors, seem to be related to the low-mass PMS objects (T Tauri stars), which have massive circumstellar disks. 
Their outbursts are generally attributed to the infall of material from the circumstellar disk to the central star (Hartmann \& Kenyon 1996).

The large-amplitude outburst of V2493 Cyg (HBC 722) in the summer of 2010 (Semkov \& Peneva 2010a, Miller et al. 2011) generated considerable interest and was studied across a wide spectral range.
Before the outburst, the target had not been registered as a variable and there were no published data from its optical photometric studies. 
Follow-up photometric and spectroscopic observations by Semkov \& Peneva (2010b), Munari et al. (2010), Leoni et al. (2010), Semkov et al. (2010), and K{\'o}sp{\'a}l et al. (2011) recorded an ongoing light increase in both the optical and infrared, as well as significant changes in the spectrum of V2493 Cyg.
Miller et al. (2011) reported on their infrared photometry and spectroscopy, as well as low and high resolution optical spectroscopy of the target. They confirmed its FUor nature and noted that it is the first FUor associated with a previously known young stellar object.
The authors also commented on the spectral classification of V2493 Cyg appearing to progress from G-type at blue optical wavelengths to K-type toward the red. 

Results from a high resolution, optical, spectroscopic campaign of V2493 Cyg during November-December 2010 are given by Lee et al. (2011). 
Optical and near-infrared $JHK$ photometry of V2493 Cyg was presented by Lorenzetti et al. (2011), K{\'o}sp{\'a}l et al. (2011), and Lorenzetti et al. (2012). 
The pre-outburst spectral energy distribution (SED) of V2493 Cyg is discussed in the papers of Miller et al. (2011) and K{\'o}sp{\'a}l et al. (2011). 
These authors concluded that V2493 Cyg had been a Class II young stellar object before the eruption, which is an object type most often associated with Classical T Tauri stars. 
The pre-outburst bolometric luminosity of the object was estimated as 0.85 $L_{\sun}$ (K{\'o}sp{\'a}l et al. 2011) while during the maximum light this luminosity rose to $\sim$12$L_{\sun}$ (Miller et al. 2011).

We also highlight the case of V2493 Cyg because it is the first (and so far the only) FUor object detected in X-rays during outburst. 
The detection was obtained by Pooley \& Green (2010) observing in the 0.2-10 keV band with the Swift X-ray satellite.
Green et al. (2011) analyzed the submillimeter CO emission surrounding V2493 Cyg using images and spectroscopy from the Herschel Space Observatory and the Caltech Submillimeter Observatory. 
The authors concluded that V2493 Cyg ``did not show evidence for a circumstellar envelope or shocked gas, and appeared to erupt from a disk-like state, similar to FU Orionis itself''.

\section{Observations}

\subsection{Photometric CCD observations}
 
The CCD photometric observations of V2493 Cyg were performed with the 2-m RCC, the 50/70-cm Schmidt, and the 60-cm Cassegrain telescopes of the National Astronomical Observatory Rozhen (Bulgaria), the 1.3-m RC telescope of the Skinakas Observatory\footnote{Skinakas Observatory is a collaborative
project of the University of Crete, the Foundation for Research and Technology - Hellas, and the Max-Planck-Institut f\"{u}r Extraterrestrische
Physik.} of the Institute of Astronomy, University of Crete (Greece), the 42-cm telescope of the Galaxy View Observatory, Sequim (USA), and both the 25-cm Newtonian and the 29-cm Schmidt Cassegrain telescopes of the Observatorio Astron\'{o}mico del CIECEM, Universidad de Huelva (Spain).  
All frames were exposed through a set of standard Johnson-Cousins filters.  

Our first CCD observations of the field around V2493 Cyg were performed with the Rozhen 2-m RCC telescope and SBIG ST-6 camera in 1994-1996. 
The subsequent set of observations with this telescope were made with the CCD cameras Photometrics AT-200 and VersArray 1300B. The observations with the 50/70-cm Schmidt telescope were performed with the cameras: SBIG ST-8 in 2000-2007, SBIG STL11000 from 2008 to Apr. 16, 2009, and FLI PL 16803 from July 2009 to the present.
Observations with the Skinakas 1.3-m RC telescope were performed with the Potometrics CH-360 CCD camera from Jun. 2000 to Jul. 3, 2007 and with the ANDOR DZ 436 CCD camera from Jul. 23, 2007 to the present.
After registration of the outburst, observations of V2493 Cyg were made with: the 60-cm Cassegrain telescopes of the Rozhen Observatory with the FLI PL 9000 CCD camera, the 42-cm telescope of the Galaxy View Observatory with the SBIG ST-8XME CCD camera, and the 25-cm Newtonian and the 29-cm Schmidt Cassegrain telescopes of the Observatorio del CIECEM with a QSI-516ws CCD camera.

All the data were analyzed using the same aperture, which was chosen as 4\arcsec in radius (while the background annulus was from 13\arcsec to 19\arcsec) in order to minimize the light from the surrounding nebula and avoid contamination from adjacent stars. 
As references, we used the $BVRI$ comparison sequence of fifteen stars in the field around V2493 Cyg published in Semkov et al. (2010). 
In this way we provided a maximum consistency of the photometric results obtained at different telescopes and CCD cameras.

The results of our photometric CCD observations of V2493 Cyg are summarized in Table 1.  
The columns provide the date and Julian date (JD) of observation, $\it IRVB $ magnitudes of V2493 Cyg, the telescope and CCD camera used. 
The typical errors in the reported magnitudes before the outburst are $0\fm01$-$0\fm02$ for $I$ and $R$-band data, $0\fm02$-$0\fm06$ for $V$, and $0\fm03-0\fm10$ for $B$-band.
The large increase in the brightness associated with the outburst contributed to a significant reduction in the error in the CCD photometry of V2493 Cyg obtained during the ongoing bright state.
Since August 2010 the typical errors in the reported magnitudes are $0\fm01$ for $I$ and $R$-band data, $0\fm01$-$0\fm02$ for $V$, and $0\fm02-0\fm05$ for $B$-band.

\onllongtab{1}{
\begin{longtable}{llllllll}
\caption{Photometric CCD observations of V2493 Cyg during the period 1994-2012.}\\
\hline
\hline
\noalign{\smallskip}
Date &	J.D. (24...)	&	I	&	R	&	V	&	B & Telescope & CCD	\\
\noalign{\smallskip}
\hline
\endfirsthead
\caption{Continued.}\\ 
\hline
\hline
\noalign{\smallskip}
Date &	J.D.(24...)	&	I	&	R	&	V	&	B & Telescope & CCD	\\
\noalign{\smallskip}
\hline
\endhead
\hline
\endfoot
\hline
\noalign{\smallskip}
\endlastfoot
1994 Aug 08	& 49573.173	& 15.31	& 16.74	& 18.32	&	-     & 2m	  & ST6 \\
1995 Nov 25	& 50046.917	& 15.23	& 16.69	& 18.27	&	-     & 2m	  & ST6 \\
1996 Jul 08 & 50272.623 & 15.14 & 16.56 & 17.92 & -     & 2m    & ST6 \\
1997 Jun 01	&	50601.491	&	15.27	&	16.85	&	18.15	&	20.0	&	2m	  &	Phot	\\		
2000 Jun 14	&	51710.467	&	15.15	&	-	    &	18.04	&	-	    &	1.3m	&	Phot	\\		
2000 Jun 16	&	51711.507	&	15.16	&	-	    &	18.08	&	-    	&	1.3m	&	Phot	\\		
2000 Jun 17	&	51712.510	&	15.16	&	-	    &	18.07	&	19.65	&	1.3m	&	Phot	\\		
2000 Jun 20	&	51716.377	&	15.12	&	-	    &	17.97	&	-   	&	1.3m	&	Phot	\\		
2000 Jun 22	&	51717.538	&	15.19	&	16.68	&	18.02	&	-   	&	1.3m	&	Phot	\\		
2000 Jun 23	&	51718.545	&	15.14	&	-   	&	18.08	&	-   	&	1.3m	&	Phot	\\		
2000 Jun 23	&	51719.470	&	15.16	&	16.71	&	18.09	&	19.77	&	1.3m	&	Phot	\\		
2000 Jun 25	&	51720.515	&	15.12	&	16.66	&	18.00	&	19.61	&	1.3m	&	Phot	\\		
2000 Oct 29	&	51847.241	&	15.07	&	16.73	&	18.01	&	-   	&	Sch	  &	ST8	\\		
2000 Oct 30	&	51848.300	&	15.08	&	16.73	&	18.01	&	-   	&	Sch 	&	ST8	\\		
2000 Dec 24	&	51903.243	&	15.07	&	16.70	&	17.93	&	-   	&	Sch	  &	ST8	\\		
2001 May 27	&	52057.425	&	15.16	&	16.97	&	18.29	&	-   	&	Sch 	&	ST8	\\		
2001 Jul 15	&	52106.549	&	15.15	&	-   	&	18.01	&	19.58	&	1.3m	&	Phot	\\		
2001 Aug 06	&	52128.486	&	15.04	&	16.55	&	17.86	&	19.52	&	1.3m	&	Phot	\\		
2001 Sep 02	&	52154.515	&	15.22	&	-   	&	18.25	&	19.56	&	1.3m	&	Phot	\\
2002 Jun 24	&	52449.000	&	15.22	&	-   	&	18.27	&	19.69	&	1.3m	&	Phot	\\
2002 Jul 07	&	52433.445	&	15.17	&	16.81	&	18.12	&	19.67	&	1.3m	&	Phot	\\
2002 Jul 15	&	52471.468	&	15.12	&	-   	&	18.17	&	-   	&	1.3m	&	Phot	\\
2002 Oct 03	&	52551.364	&	15.17	&	16.84	&	18.00	&	-   	&	Sch	  &	ST8	\\
2002 Oct 04	&	52552.394	&	15.27	&	17.12	&	18.43	&	-   	&	Sch	  &	ST8	\\
2002 Oct 29	&	52577.288	&	15.25	&	17.09	&	18.50	&	-   	&	Sch 	&	ST8	\\
2002 Oct 30	&	52578.284	&	15.16	&	17.00	&	18.26	&	-   	&	Sch	  &	ST8	\\
2002 Nov 01	&	52580.193	&	15.10	&	16.82	&	18.01	&	-   	&	Sch	  &	ST8	\\
2002 Nov 28	&	52607.212	&	15.23	&	16.98	&	18.22	&	-   	&	Sch	  &	ST8	\\
2003 Mar 03	&	52701.593	&	15.25	&	17.02	&	18.28	&	19.41	&	2m	  &	Phot	\\
2003 May 06	&	52765.517	&	15.05	&	16.70	&	18.08	&	-   	&	Sch	  &	ST8	\\
2003 Sep 27	&	52910.375	&	15.21	&	17.05	&	18.34	&	-   	&	Sch	  &	ST8	\\
2003 Nov 25	&	52969.192	&	15.08	&	16.77	&	18.00	&	-   	&	Sch	  &	ST8	\\
2004 Mar 22	&	53086.582	&	15.09	&	16.76	&	18.11	&	-   	&	2m  	&	Phot	\\
2004 Jul 17	&	53203.607	&	15.14	&	16.79	&	18.19	&	-   	&	Sch 	&	ST8	\\
2004 Aug 20	&	53238.328	&	15.09	&	16.64	&	17.96	&	19.53	&	1.3m	&	Phot	\\
2004 Sep 08	&	53257.324	&	14.99	&	16.46	&	17.73	&	19.27	&	1.3m	&	Phot	\\
2004 Sep 09	&	53258.408	&	15.11	&	16.71	&	18.12	&	19.71	&	1.3m	&	Phot	\\
2004 Sep 28	&	53277.227	&	15.01	&	16.47	&	17.81	&	19.27	&	1.3m	&	Phot	\\
2004 Sep 29	&	52278.237	&	15.06	&	16.60	&	17.86	&	19.49	&	1.3m	&	Phot	\\
2004 Sep 30	&	53279.274	&	14.99	&	16.49	&	17.75	&	19.33	&	1.3m	&	Phot	\\
2004 Nov 18	&	53328.231	&	15.07	&	16.69	&	17.83	&	-   	&	Sch  	&	ST8	\\
2004 Nov 20	&	53330.273	&	15.00	&	16.56	&	17.77	&	-   	&	Sch	  &	ST8	\\
2005 Aug 23	&	53606.289	&	15.11	&	16.78	&	18.15	&	-     &	1.3m	&	Phot	\\
2005 Aug 25	&	53609.509	&	15.20	&	16.89	&	18.26	&	-   	&	1.3m	&	Phot	\\
2005 Aug 27	&	53610.438	&	15.17	&	16.87	&	18.27	&	19.84	&	1.3m	&	Phot	\\
2005 Sep 14	&	53628.281	&	15.16	&	16.76	&	18.09	&	-   	&	1.3m	&	Phot	\\
2006 Mar 28	&	53821.545	&	15.06	&	16.71	&	18.00	&	19.70	&	2m  	&	VA	\\
2006 Jul 19	&	53936.474	&	15.11	&	16.81	&	18.16	&	-   	&	Sch	  &	ST8	\\
2006 Jul 21	&	53938.394	&	15.09	&	-   	&	17.99	&	19.58	&	2m	  &	Phot	\\
2006 Sep 30	&	54009.385	&	15.16	&	16.85	&	18.23	&	19.76	&	1.3m	&	Phot	\\
2006 Oct 05	&	54014.359	&	15.14	&	16.81	&	18.19	&	19.66	&	1.3m	&	Phot	\\
2006 Dec 16	&	54086.204	&	15.18	&	16.74	&	18.20	&	-   	&	Sch 	&	ST8	\\
2007 Jun 26	&	54278.380	&	15.09	&	16.64	&	17.99	&	19.60	&	1.3m	&	Phot	\\
2007 Jul 03	&	54285.341	&	15.20	&	16.80	&	18.23	&	19.80	&	1.3m	&	Phot	\\
2007 Jul 23	&	54305.319	&	15.05	&	16.61	&	17.92	&	19.46	&	1.3m	&	ANDOR	\\
2007 Jul 24	&	54306.317	&	15.22	&	16.80	&	18.13	&	19.58	&	1.3m	&	ANDOR	\\
2007 Aug 16	&	54329.384	&	15.00	&	16.57	&	17.88	&	16.60	&	2m	  &	VA	\\
2007 Aug 17	&	54330.289	&	15.08	&	16.53	&	17.90	&	19.46	&	2m	  &	VA	\\
2008 Jun 28	&	54646.366	&	15.21	&	16.91	&	18.29	&	19.82	&	1.3m	&	ANDOR	\\
2008 Jun 29	&	54647.384	&	15.16	&	16.88	&	18.24	&	19.86	&	1.3m	&	ANDOR	\\
2008 Jul 05	&	54653.347	&	15.23	&	16.95	&	18.36	&	20.04	&	1.3m	&	ANDOR	\\
2008 Jul 06	&	54654.376	&	15.28	&	16.96	&	18.29	&	19.90	&	1.3m	&	ANDOR	\\
2008 Jul 25	&	54673.331	&	15.26	&	16.96	&	18.37	&	19.88	&	1.3m	&	ANDOR	\\
2008 Aug 28	&	54707.312	&	15.30	&	16.81	&	18.22	&	-   	&	Sch 	&	STL11000	\\
2008 Oct 23	&	54763.202	&	15.26	&	16.86	&	18.21	&	-   	&	Sch	  &	STL11000	\\
2009 Apr 16	&	54938.551	&	15.24	&	16.80	&	18.12	&	-   	&	Sch 	&	STL11000	\\
2009 Jun 17	&	55000.499	&	15.23	&	17.02	&	18.42	&	19.80	&	1.3m	&	ANDOR	\\
2009 Jun 27	&	55009.522	&	15.22	&	16.86	&	18.23	&	19.74	&	1.3m	&	ANDOR	\\
2009 Jun 28	&	55011.458	&	15.31	&	16.99	&	18.32	&	-   	&	Sch 	&	FLI	\\
2009 Jul 07	&	55016.501	&	15.28	&	16.93	&	18.33	&	19.97	&	1.3m	&	ANDOR	\\
2009 Jul 10	&	55022.502	&	15.24	&	17.01	&	18.42	&	19.82	&	1.3m	&	ANDOR	\\
2009 Jul 14	&	55027.412	&	15.33	&	17.10	&	18.40	&	-   	&	Sch	  &	FLI	\\
2009 Jul 15	&	55028.380	&	15.19	&	16.91	&	-   	&	-   	&	Sch	  &	FLI	\\
2009 Jul 23	&	55035.502	&	15.18	&	16.82	&	18.18	&	19.80	&	1.3m	&	ANDOR	\\
2009 Jul 31	&	55044.338	&	15.28	&	16.98	&	18.33	&	19.90	&	1.3m	&	ANDOR	\\
2009 Aug 21	&	55065.292	&	15.17	&	16.88	&	18.35	&	-   	&	Sch	  &	FLI	\\
2009 Oct 06	&	55111.330	&	15.29	&	-   	&	18.45	&	-   	&	Sch	  &	FLI	\\
2009 Oct 08	&	55113.243	&	15.30	&	17.10	&	18.38	&	-   	&	Sch 	&	FLI	\\
2009 Nov 20	&	55156.197	&	15.25	&	17.00	&	18.47	&	-   	&	Sch 	&	FLI	\\
2009 Nov 21	&	55157.227	&	15.19	&	16.90	&	18.25	&	-   	&	Sch	  &	FLI	\\
2010 May 13	&	55330.436	&	14.60	&	16.04	&	17.37	&	18.96	&	Sch	  &	FLI	\\
2010 Jun 10	&	55358.424	&	14.53	&	16.06	&	17.26	&	18.89	&	Sch	  &	FLI	\\
2010 Aug 06	&	55415.466	&	12.38	&	13.53	&	14.59	&	16.10	&	Sch	  &	FLI	\\
2010 Aug 07	&	55416.433	&	12.32	&	13.45	&	14.51	&	16.04	&	Sch	  &	FLI	\\
2010 Aug 11	&	55420.497	&	12.12	&	13.25	&	14.30	&	15.84	&	1.3m	&	ANDOR	\\
2010 Aug 12	&	55421.326	&	12.11	&	13.24	&	14.29	&	15.84	&	1.3m	&	ANDOR	\\
2010 Aug 13	&	55422.258	&	12.07	&	13.20	&	14.24	&	15.76	&	1.3m	&	ANDOR	\\
2010 Aug 14	&	55423.256	&	12.04	&	13.14	&	14.18	&	15.70	&	1.3m	&	ANDOR	\\
2010 Aug 15	&	55424.251	&	12.00	&	13.10	&	14.15	&	15.68	&	1.3m	&	ANDOR	\\
2010 Aug 16	&	55425.250	&	11.94	&	13.04	&	14.07	&	15.61	&	1.3m	&	ANDOR	\\
2010 Aug 18	&	55426.599	&	11.89	&	12.99	&	14.03	&	15.54	&	1.3m	&	ANDOR	\\
2010 Aug 19	&	55427.551	&	-   	&	12.94	&	13.99	&	-   	&	0.25m	&	QSI	\\
2010 Aug 19	&	55427.591	&	11.87	&	12.97	&	14.02	&	15.54	&	1.3m	&	ANDOR	\\
2010 Aug 20	&	55428.590	&	11.88	&	13.00	&	14.05	&	15.57	&	1.3m	&	ANDOR	\\
2010 Aug 21	&	55429.589	&	11.88	&	12.99	&	14.05	&	15.58	&	1.3m	&	ANDOR	\\
2010 Aug 21	&	55429.665	&	-   	&	12.97	&	14.01	&	-   	&	0.25m	&	QSI	\\
2010 Aug 22	&	55430.665	&	-   	&	12.91	&	13.93	&	-   	&	0.25m	&	QSI	\\
2010 Aug 23	&	55431.665	&	-   	&	12.85	&	13.88	&	-   	&	0.25m	&	QSI	\\
2010 Aug 23	&	55431.747	&	11.81	&	12.83	&	13.86	&	15.51	&	0.42m	&	ST8	\\
2010 Aug 23	&	55432.519	&	11.75	&	12.83	&	13.86	&	15.37	&	1.3m	&	ANDOR	\\
2010 Aug 24	&	55432.662	&	-   	&	12.80	&	13.80	&	-   	&	0.25m	&	QSI	\\
2010 Aug 24	&	55433.470	&	11.71	&	12.79	&	13.81	&	15.32	&	1.3m	&	ANDOR	\\
2010 Aug 25	&	55433.651	&	-   	&	12.75	&	13.77	&	-   	&	0.25m	&	QSI	\\
2010 Aug 25	&	55433.758	&	11.74	&	12.74	&	13.77	&	15.31	&	0.42m	&	ST8	\\
2010 Aug 25	&	55434.290	&	11.72	&	12.81	&	13.83	&	15.34	&	1.3m	&	ANDOR	\\
2010 Aug 26	&	55434.528	&	-   	&	12.80	&	13.83	&	-   	&	0.25m	&	QSI	\\
2010 Aug 26	&	55435.310	&	11.74	&	12.83	&	13.84	&	15.36	&	1.3m	&	ANDOR	\\
2010 Aug 27	&	55435.501	&	-   	&	12.79	&	13.83	&	-   	&	0.25m	&	QSI	\\
2010 Aug 27	&	55436.282	&	11.74	&	12.84	&	13.85	&	15.36	&	60cm	&	FLI	\\
2010 Aug 28	&	55436.594	&	-   	&	12.78	&	13.80	&	-   	&	0.25m	&	QSI	\\
2010 Aug 28	&	55437.282	&	11.76	&	12.84	&	13.87	&	15.33	&	60cm	&	FLI	\\
2010 Aug 29	&	55437.621	&	-   	&	12.80	&	13.83	&	-   	&	0.25m	&	QSI	\\
2010 Aug 29	&	55437.718	&	11.78	&	12.76	&	13.83	&	15.36	&	0.42m	&	ST8	\\
2010 Aug 29	&	55438.456	&	-   	&	12.76	&	13.78	&	-   	&	0.25m	&	QSI	\\
2010 Aug 30	&	55439.237	&	11.70	&	12.78	&	13.82	&	15.33	&	1.3m	&	ANDOR	\\
2010 Aug 31	&	55440.238	&	11.68	&	12.75	&	13.77	&	15.28	&	1.3m	&	ANDOR	\\
2010 Sep 01	&	55441.361	&	-   	&	12.70	&	13.71	&	-   	&	0.25m	&	QSI	\\
2010 Sep 02	&	55442.269	&	11.68	&	12.81	&	13.78	&	15.31	&	60cm	&	FLI	\\
2010 Sep 03	&	55443.496	&	-   	&	12.71	&	13.74	&	-   	&	0.25m	&	QSI	\\
2010 Sep 08	&	55448.326	&	11.63	&	12.70	&	13.68	&	15.16	&	Sch 	&	FLI	\\
2010 Sep 07	&	55447.420	&	11.62	&	12.70	&	13.69	&	15.17	&	Sch	  &	FLI	\\
2010 Sep 09	&	55449.407	&	11.64	&	12.73	&	13.72	&	15.20	&	Sch	  &	FLI	\\
2010 Sep 13	&	55453.351	&	-   	&	12.63	&	13.63	&	-   	&	0.25m	&	QSI	\\
2010 Sep 14	&	55453.702	&	11.62	&	12.64	&	13.67	&	15.22	&	0.42m	&	ST8	\\
2010 Sep 14	&	55454.347	&	-   	&	12.64	&	13.63	&	-   	&	0.25m	&	QSI	\\
2010 Sep 15	&	55454.744	&	11.60	&	12.58	&	13.63	&	15.19	&	0.42m	&	ST8	\\
2010 Sep 17	&	55457.399	&	-   	&	12.61	&	13.61	&	-   	&	0.25m	&	QSI	\\
2010 Sep 18	&	55458.218	&	11.59	&	12.65	&	13.65	&	15.15	&	1.3m	&	ANDOR	\\
2010 Sep 18	&	55458.260	&	11.59	&	12.64	&	13.65	&	15.15	&	60cm	&	FLI	\\
2010 Sep 18	&	55458.348	&	-   	&	12.61	&	13.62	&	-   	&	0.25m	&	QSI	\\
2010 Sep 19	&	55459.345	&	-   	&	12.64	&	13.67	&	-   	&	0.25m	&	QSI	\\
2010 Sep 20	&	55459.513	&	11.61	&	12.67	&	13.69	&	15.19	&	1.3m	&	ANDOR	\\
2010 Sep 20	&	55460.488	&	11.65	&	12.69	&	13.75	&	-   	&	60cm	&	FLI	\\
2010 Sep 21	&	55460.737	&	11.65	&	12.67	&	13.69	&	15.22	&	0.42m	&	ST8	\\
2010 Sep 21	&	55461.342	&	-   	&	12.70	&	13.73	&	-   	&	0.25m	&	QSI	\\
2010 Sep 22	&	55461.707	&	11.65	&	12.65	&	13.71	&	15.25	&	0.42m	&	ST8	\\
2010 Sep 26	&	55465.645	&	11.60	&	12.62	&	13.63	&	15.26	&	0.42m	&	ST8	\\
2010 Sep 26	&	55466.329	&	-   	&	12.64	&	13.67	&	-   	&	0.25m	&	QSI	\\
2010 Sep 27	&	55467.321	&	-   	&	12.66	&	13.68	&	-   	&	0.25m	&	QSI	\\
2010 Sep 28	&	55468.318	&	-   	&	12.66	&	13.68	&	-   	&	0.25m	&	QSI	\\
2010 Sep 29	&	55468.695	&	11.63	&	12.64	&	13.67	&	15.21	&	0.42m	&	ST8	\\
2010 Sep 29	&	55469.315	&	-   	&	12.64	&	13.66	&	-   	&	0.25m	&	QSI	\\
2010 Sep 30	&	55469.721	&	11.62	&	12.61	&	13.67	&	15.18	&	0.42m	&	ST8	\\
2010 Oct 01	&	55470.718	&	11.61	&	12.61	&	13.65	&	15.18	&	0.42m	&	ST8	\\
2010 Oct 01	&	55471.334	&	-   	&	12.62	&	13.65	&	-   	&	0.25m	&	QSI	\\
2010 Oct 02	&	55472.314	&	-   	&	12.66	&	13.64	&	-   	&	0.25m	&	QSI	\\
2010 Oct 04	&	55474.467	&	-   	&	12.68	&	13.70	&	-   	&	0.25m	&	QSI	\\
2010 Oct 05	&	55475.312	&	-   	&	12.72	&	13.74	&	-   	&	0.25m	&	QSI	\\
2010 Oct 06	&	55476.315	&	-   	&	12.73	&	13.76	&	-   	&	0.25m	&	QSI	\\
2010 Oct 07	&	55476.679	&	11.68	&	12.69	&	13.73	&	15.24	&	0.42m	&	ST8	\\
2010 Oct 11	&	55480.680	&	11.68	&	12.71	&	13.75	&	15.29	&	0.42m	&	ST8	\\
2010 Oct 11	&	55481.276	&	11.59	&	12.68	&	13.69	&	15.18	&	1.3m	&	ANDOR	\\
2010 Oct 14	&	55483.544	&	-   	&	12.75	&	13.76	&	-   	&	0.25m	&	QSI	\\
2010 Oct 14	&	55483.738	&	11.72	&	12.74	&	13.77	&	15.30	&	0.42m	&	ST8	\\
2010 Oct 15	&	55484.505	&	-   	&	12.73	&	13.74	&	-   	&	0.25m	&	QSI	\\
2010 Oct 16	&	55485.507	&	-   	&	12.81	&	13.84	&	-   	&	0.25m	&	QSI	\\
2010 Oct 17	&	55486.504	&	-   	&	12.79	&	13.82	&	-   	&	0.25m	&	QSI	\\
2010 Oct 17	&	55486.695	&	11.76	&	12.77	&	13.82	&	15.33	&	0.42m	&	ST8	\\
2010 Oct 18	&	55487.504	&	-   	&	12.83	&	13.85	&	-   	&	0.25m	&	QSI	\\
2010 Oct 20	&	55489.686	&	11.80	&	12.83	&	13.88	&	15.44	&	0.42m	&	ST8	\\
2010 Oct 21	&	55490.686	&	11.79	&	12.82	&	13.86	&	15.39	&	0.42m	&	ST8	\\
2010 Oct 21	&	55491.297	&	-   	&	12.82	&	13.85	&	-   	&	0.25m	&	QSI	\\
2010 Oct 22	&	55492.203	&	11.77	&	12.87	&	13.89	&	15.42	&	60cm	&	FLI	\\
2010 Oct 23	&	55493.216	&	11.82	&	12.93	&	13.97	&	15.45	&	60cm	&	FLI	\\
2010 Oct 23	&	55493.331	&	-   	&	12.89	&	13.91	&	-   	&	0.25m	&	QSI	\\
2010 Oct 24	&	55494.216	&	11.84	&	12.96	&	13.97	&	-   	&	60cm	&	FLI	\\
2010 Oct 29	&	55499.257	&	11.88	&	-   	&	13.95	&	15.51	&	2m	  &	VA	\\
2010 Oct 30	&	55500.215	&	11.87	&	-   	&	13.93	&	15.49	&	2m	  &	VA	\\
2010 Oct 31	&	55501.195	&	11.85	&	12.97	&	13.99	&	15.45	&	Sch	  &	FLI	\\
2010 Oct 31	&	55501.287	&	11.85	&	-   	&	13.93	&	15.48	&	2m	  &	VA	\\
2010 Nov 01	&	55502.234	&	11.87	&	-   	&	13.95	&	15.50	&	2m	  &	VA	\\
2010 Nov 02	&	55503.223	&	11.87	&	13.00	&	14.03	&	15.52	&	Sch  	&	FLI	\\
2010 Nov 03	&	55503.699	&	11.88	&	12.91	&	13.96	&	15.51	&	0.42m	&	ST8	\\
2010 Nov 03	&	55504.195	&	11.87	&	12.98	&	14.00	&	15.48	&	Sch	  &	FLI	\\
2010 Nov 04	&	55504.665	&	11.89	&	12.93	&	13.94	&	15.47	&	0.42m	&	ST8	\\
2010 Nov 04	&	55505.202	&	11.86	&	12.97	&	13.98	&	15.46	&	Sch	  &	FLI	\\
2010 Nov 05	&	55506.226	&	11.86	&	12.97	&	13.98	&	15.47	&	Sch	  &	FLI	\\
2010 Nov 06	&	55507.224	&	11.85	&	12.96	&	13.98	&	15.47	&	Sch	  &	FLI	\\
2010 Nov 12	&	55512.725	&	11.94	&	12.96	&	13.98	&	15.57	&	0.42m	&	ST8	\\
2010 Nov 13	&	55514.241	&	11.95	&	13.05	&	14.08	&	15.56	&	60cm	&	FLI	\\
2010 Nov 14	&	55515.217	&	11.95	&	13.05	&	14.08	&	15.53	&	60cm	&	FLI	\\
2010 Nov 16	&	55517.269	&	12.00	&	13.12	&	14.16	&	15.57	&	60cm	&	FLI	\\
2010 Nov 21	&	55522.222	&	12.00	&	13.14	&	14.16	&	15.68	&	Sch	  &	FLI	\\
2010 Nov 25	&	55526.206	&	12.03	&	13.16	&	14.18	&	15.68	&	Sch	  &	FLI	\\
2010 Nov 27	&	55528.266	&	12.08	&	13.20	&	14.22	&	15.69	&	Sch	  &	FLI	\\
2010 Nov 29	&	55529.642	&	12.09	&	13.16	&	14.19	&	15.79	&	0.42m	&	ST8	\\
2010 Dec 03	&	55533.741	&	12.13	&	13.20	&	14.25	&	15.72	&	0.42m	&	ST8	\\
2010 Dec 05	&	55535.612	&	12.16	&	13.22	&	14.28	&	15.71	&	0.42m	&	ST8	\\
2010 Dec 06	&	55536.619	&	12.16	&	13.23	&	14.28	&	15.63	&	0.42m	&	ST8	\\
2010 Dec 07	&	55538.251	&	12.15	&	13.29	&	14.31	&	15.80	&	60cm	&	FLI	\\
2010 Dec 23	&	55554.289	&	-   	&	13.38	&	14.45	&	-   	&	0.25m	&	QSI	\\
2010 Dec 31	&	55561.615	&	12.33	&	13.43	&	14.49	&	16.01	&	0.42m	&	ST8	\\
2011 Jan 01	&	55562.599	&	12.39	&	13.50	&	14.56	&	16.09	&	0.42m	&	ST8	\\
2011 Jan 01	&	55563.169	&	12.39	&	13.57	&	14.59	&	16.08	&	Sch	  &	FLI	\\
2011 Jan 03	&	55564.625	&	12.38	&	13.48	&	14.55	&	16.10	&	0.42m	&	ST8	\\
2011 Jan 04	&	55566.282	&	-   	&	13.53	&	14.55	&	-   	&	0.25m	&	QSI	\\
2011 Jan 06	&	55568.198	&	12.45	&	-   	&	14.68	&	16.22	&	2m  	&	VA	\\
2011 Jan 07	&	55569.210	&	12.43	&	-   	&	14.66	&	-   	&	2m	  &	VA	\\
2011 Jan 08	&	55570.204	&	12.49	&	-   	&	14.69	&	16.22	&	2m  	&	VA	\\
2011 Jan 09	&	55570.617	&	12.45	&	13.56	&	14.63	&	16.17	&	0.42m	&	ST8	\\
2011 Jan 09	&	55571.216	&	12.44	&	-   	&	14.67	&	16.20	&	2m  	&	VA	\\
2011 Jan 10	&	55571.625	&	12.48	&	13.59	&	14.66	&	16.26	&	0.42m	&	ST8	\\
2011 Jan 10	&	55572.175	&	12.41	&	-    	&	-   	&	-   	&	2m	  &	VA	\\
2011 Jan 10	&	55572.278	&	-   	&	13.59	&	14.67	&	-   	&	0.25m	&	QSI	\\
2011 Jan 11	&	55573.232	&	12.49	&	-   	&	14.69	&	16.24	&	2m  	&	VA	\\
2011 Jan 12	&	55574.219	&	12.48	&	-   	&	14.69	&	16.23	&	2m  	&	VA	\\
2011 Jan 13	&	55575.288	&	-   	&	13.66	&	14.71	&	-   	&	0.25m	&	QSI	\\
2011 Feb 02	&	55594.624	&	12.54	&	13.66	&	14.71	&	16.27	&	0.42m	&	ST8	\\
2011 Feb 06	&	55599.190	&	12.61	&	13.79	&	14.83	&	-   	&	Sch	  &	FLI	\\
2011 Feb 07	&	55600.202	&	12.57	&	13.80	&	14.85	&	-   	&	Sch	  &	FLI	\\
2011 Feb 08	&	55600.671	&	12.54	&	13.76	&	14.81	&	-   	&	Sch	  &	FLI	\\
2011 Feb 26	&	55619.647	&	12.55	&	13.76	&	14.82	&	16.39	&	Sch	  &	FLI	\\
2011 Mar 15	&	55635.640	&	12.62	&	13.86	&	14.92	&	16.40	&	Sch	  &	FLI	\\
2011 Mar 16	&	55636.646	&	12.62	&	13.84	&	14.92	&	16.45	&	Sch	  &	FLI	\\
2011 Mar 18	&	55638.716	&	-   	&	13.90	&	14.99	&	-   	&	0.29m	&	QSI	\\
2011 Mar 19	&	55639.715	&	-   	&	13.86	&	14.98	&	-   	&	0.29m	&	QSI	\\
2011 Mar 24	&	55644.716	&	-   	&	13.93	&	15.02	&	-   	&	0.29m	&	QSI	\\
2011 Mar 27	&	55648.632	&	12.67	&	13.89	&	14.98	&	16.50	&	Sch	  &	FLI	\\
2011 Mar 30	&	55650.708	&	-   	&	13.90	&	14.98	&	-   	&	0.29m	&	QSI	\\
2011 Mar 31	&	55651.707	&	-   	&	13.90	&	15.00	&	-   	&	0.29m	&	QSI	\\
2011 Apr 01	&	55652.706	&	-   	&	13.89	&	14.98	&	-   	&	0.29m	&	QSI	\\
2011 Apr 04	&	55656.509	&	12.64	&	13.86	&	14.97	&	16.50	&	Sch	  &	FLI	\\
2011 Apr 05	&	55656.706	&	-   	&	13.87	&	15.00	&	-   	&	0.29m	&	QSI	\\
2011 Apr 07	&	55659.519	&	12.67	&	-   	&	14.95	&	16.54	&	2m	  &	VA	\\
2011 Apr 10	&	55661.697	&	-   	&	13.91	&	14.98	&	-   	&	0.29m	&	QSI	\\
2011 Apr 11	&	55662.697	&	-   	&	13.88	&	14.96	&	-   	&	0.29m	&	QSI	\\
2011 Apr 11	&	55663.619	&	12.65	&	13.89	&	15.00	&	16.53	&	Sch  	&	FLI	\\
2011 Apr 14	&	55666.619	&	12.63	&	13.89	&	15.01	&	-   	&	Sch	  &	FLI	\\
2011 Apr 16	&	55667.691	&	-   	&	13.89	&	14.94	&	-   	&	0.29m	&	QSI	\\
2011 Apr 26	&	55677.671	&	-   	&	13.96	&	15.04	&	-   	&	0.29m	&	QSI	\\
2011 Apr 28	&	55679.671	&	-   	&	13.95	&	15.05	&	-   	&	0.29m	&	QSI	\\
2011 May 01	&	55683.573	&	12.68	&	13.93	&	15.02	&	16.63	&	2m  	&	VA	\\
2011 May 04	&	55685.648	&	-   	&	13.98	&	15.09	&	-   	&	0.29m	&	QSI	\\
2011 May 21	&	55703.384	&	12.64	&	13.89	&	14.98	&	16.55	&	Sch	  &	FLI	\\
2011 May 22	&	55704.430	&	12.68	&	13.96	&	15.05	&	16.73	&	Sch	  &	FLI	\\
2011 May 23	&	55705.408	&	12.64	&	13.91	&	15.01	&	16.59	&	Sch	  &	FLI	\\
2011 May 24	&	55706.382	&	12.68	&	13.96	&	15.09	&	16.66	&	Sch	  &	FLI	\\
2011 May 25	&	55707.390	&	12.68	&	13.91	&	15.03	&	16.57	&	Sch	  &	FLI	\\
2011 Jun 08	&	55721.380	&	12.67	&	13.83	&	15.04	&	16.67	&	2m	  &	VA	\\
2011 Jun 09	&	55722.425	&	12.60	&	13.89	&	15.00	&	16.56	&	Sch	  &	FLI	\\
2011 Jun 21	&	55734.420	&	12.55	&	13.82	&	14.96	&	16.53	&	Sch	  &	FLI	\\
2011 Jun 22	&	55735.452	&	12.52	&	13.81	&	14.92	&	16.50	&	Sch	  &	FLI	\\
2011 Jun 23	&	55736.443	&	12.55	&	13.87	&	14.97	&	16.57	&	Sch	  &	FLI	\\
2011 Jun 24	&	55737.443	&	12.53	&	13.81	&	14.94	&	16.52	&	Sch	  &	FLI	\\
2011 Jun 26	&	55739.536	&	12.45	&	13.68	&	14.81	&	-   	&	60cm	&	FLI	\\
2011 Jul 15	&	55758.566	&	12.51	&	13.75	&	14.88	&	16.38	&	60cm	&	FLI	\\
2011 Jul 18	&	55761.568	&	12.52	&	13.77	&	14.89	&	16.55	&	60cm	&	FLI	\\
2011 Jul 27	&	55770.361	&	12.57	&	13.88	&	15.01	&	16.62	&	Sch  	&	FLI	\\
2011 Aug 16	&	55790.386	&	12.52	&	13.82	&	14.97	&	16.58	&	1.3m	&	ANDOR	\\
2011 Aug 17	&	55791.414	&	12.50	&	13.80	&	14.94	&	16.55	&	1.3m	&	ANDOR	\\
2011 Aug 18	&	55792.397	&	12.50	&	13.81	&	14.96	&	16.57	&	1.3m	&	ANDOR	\\
2011 Aug 23	&	55797.323	&	12.49	&	13.77	&	14.88	&	16.50	&	Sch	  &	FLI	\\
2011 Aug 24	&	55798.301	&	12.51	&	13.80	&	14.92	&	16.55	&	Sch	  &	FLI	\\
2011 Aug 25	&	55799.318	&	12.51	&	13.80	&	14.94	&	16.54	&	Sch	  &	FLI	\\
2011 Sep 10	&	55815.249	&	12.59	&	13.90	&	15.05	&	16.64	&	1.3m	&	ANDOR	\\
2011 Sep 11	&	55816.379	&	12.56	&	13.86	&	15.02	&	16.61	&	1.3m	&	ANDOR	\\
2011 Sep 19	&	55824.281	&	12.54	&	13.86	&	15.03	&	16.64	&	1.3m	&	ANDOR	\\
2011 Sep 23	&	55828.249	&	12.54	&	13.85	&	14.98	&	16.59	&	Sch  	&	FLI	\\
2011 Oct 07	&	55842.269	&	12.41	&	13.69	&	14.82	&	16.43	&	1.3m	&	ANDOR	\\	
2011 Oct 13	&	55848.277	&	12.39	&	13.66	&	14.79	&	16.40	&	1.3m	&	ANDOR	\\	
2011 Oct 29	&	55864.251	&	12.33	&	13.56	&	14.64	&	16.31	&	2m  	&	VA	\\	
2011 Oct 30	&	55865.186	&	12.33	&	13.61	&	14.66	&	16.33	&	2m  	&	VA	\\	
2011 Oct 31	&	55866.282	&	12.33	&	13.55	&	14.62	&	16.28	&	2m  	&	VA	\\	
2011 Nov 26	& 55892.170	& 12.23	& 13.48	& 14.58	& 16.19	& 2m	  & VA \\
2011 Nov 27	& 55893.176	& 12.28	& 13.54	& 14.63	& 16.26	& Sch	  & FLI \\
2011 Nov 28	& 55894.175	& 12.27	& 13.53	& 14.65	& 16.24	& Sch	  & FLI \\
2011 Nov 29	& 55895.167	& 12.29	& 13.56	& 14.66	& 16.29	& Sch	  & FLI \\
2011 Nov 30	& 55896.179	& 12.23	& 13.48	& 14.60	& 16.19	& Sch	  & FLI \\
2011 Dec 29 & 55925.169 & 12.04 & 13.25 & 14.34 & 15.93 & Sch   & FLI \\
2012 Jan 01	& 55928.185	& 12.02	& 13.24	& 14.31	& 15.89	& Sch	  & FLI \\
2012 Jan 02	& 55929.198	& 12.05	& 13.30	& 14.39	& 15.95	& Sch	  & FLI \\
2012 Jan 03	& 55930.191	& 12.04	& 13.26	& 14.36	& 15.96	& Sch	  & FLI \\
2012 Jan 04	& 55931.219	& 12.00	& 13.21	& 14.29	& 15.85	& Sch	  & FLI \\
2012 Jan 13	& 55940.206	& 11.95	& 13.15	& 14.20	& -     &	Sch	  & FLI \\
2012 Jan 14	& 55941.202	& 11.95	& 13.15	& 14.21	& 15.78	& Sch	  & FLI \\
2012 Jan 15	& 55942.204	& 11.97	& 13.17	& 14.24	& 15.83	& Sch	  & FLI \\
2012 Jan 31	& 55958.188	& 11.92	& 13.07	& 14.10	& 15.66	& 2m	  & VA \\
2012 Mar 16	& 56003.563	& 11.92	& 13.10	& 14.18	& 15.74	& Sch	  & FLI \\
2012 Mar 29 & 56015.506 & 11.77 & 12.91 & 13.99 & 15.60 & 2m    & VA \\
2012 Apr 13 & 56030.508 & 11.77 & 12.93 & 14.01 & 15.58 & Sch   & FLI\\
\end{longtable}
}

\subsection{Spectral observations}

Medium-resolution, single-dispersion spectroscopy of V2493 Cyg was obtained on Nov. 21, 2011 with the AFOSC spectrograph+imager mounted on the Asiago
1.82-m reflector and equipped with a 1720 ln/mm volume phase holographic grism and a 2.1 arcsec slit oriented east-west (E-W).  
Medium and low resolution, single dispersion spectroscopy of V2493 Cyg was obtained on Nov. 30 and Dec. 1, 2011, with the Asiago 1.22-m reflector equipped with 1200 ln/mm and 300 ln/mm gratings.  
The slit was widened to 2.0 arcsec and oriented E-W.  
Absolute flux calibration was obtained by observing spectrophotometric standard stars both just before and soon after the observations of V2493 Cyg and at similar airmasses. 
Finally, a 2$\times$2 binned echelle spectrum was obtained on Jan. 13, 2012 with the REOSC echelle spectrograph mounted on the 1.82-m telescope, with a 2.2 arcsec wide slit oriented E-W. 
All data reduction was performed within IRAF, with Table 2 providing a log of spectral observations.

\begin{table*}
\caption{Journal of spectroscopic observations}
\label{table:3}
\centering
\begin{tabular}{crrccc}
\hline
\multicolumn{6}{c}{}\\
date & \multicolumn{1}{c}{UT} & \multicolumn{1}{c}{expt} && $\lambda$ range& tel.\\
 & & (sec) &  & (\AA) & \\
\multicolumn{6}{c}{}\\
2011-11-21  & 20:08  & 3600  & disp. 0.72 \AA/pix   & 6350$-$7080   &  1.82m+AFOSC\\  
2011-11-30  & 20:28  & 1800  & disp. 0.61 \AA/pix   & 5680$-$6910   &  1.22m+B\&C\\  
2011-12-01  & 19:01  & 7200  & disp. 2.31 \AA/pix   & 3700$-$7550   &  1.22m+B\&C\\  
2012-01-13  & 17:19  & 3600  & res. pow. 12000      & 4400$-$7335   &  1.82m+echelle\\  
\multicolumn{6}{c}{}\\
\hline
\end{tabular}
\end{table*}

\subsection{Archival photographic observations}

The region of NGC 7000/IC 5070 is one of the most well studied areas of star formation in our galaxy.
The field is rich in young stellar objects with different masses, H$\alpha$ emission stars, Herbig-Haro objects, flare stars from UV Ceti type, and other irregular variable young objects (Reipurth \& Schneider 2008; Armond and al. 2011). 
On the other hand, the bright North America and Pelican nebulae have been continuously attracting the interest of astrophotographers and
researchers worldwide, and as a consequence the plate archives of several observatories preserve abundant collections of plates exposed over several
decades on this region of the sky.
The collection and analysis of all these observations is very valuable for the study of the long-time variability of V2493 Cyg but it requires a very long and laborious amount of work.

In this paper, we present the result of exploring the whole photographic plate stack preserved at two observatories, the Asiago Observatory (Italy), and the National Astronomical Observatory Rozhen (Bulgaria).
The digitized plates from the Palomar Schmidt telescope, available via the website of the Space Telescope Science Institute, are also used.
Photographic observations in the field of V2493 Cyg were performed with the 67/92 cm and the 40/50 cm Schmidt telescopes at the Asiago Observatory
and with the 2-m RCC telescope and the 50/70 cm Schmidt telescope at the Rozhen Observatory.
In addition, we used several scanned plates from the 100/130 cm Schmidt telescope of the Byurakan Astrophysical Observatory (Armenia), the 30 cm Astrograph of the Hoher List Observatory (Germany), and the historic first Schmidt-type telescope (36/44 cm) mounted on the Hamburg-Bergedorf Observatory (Germany). 
These plates are available from the Sofia Sky Archive Center - the Wide-field Plate Database\footnote{www.wfpdb.org}.  
The plates from Rozhen, Byurakan, Hoher List, and Hamburg-Bergedorf were scanned with Epson Expression 1640 XL/10000XL scanners, which have 1600-2400 dpi resolution and a corresponding pixel size from 16$\times$16 $\mu$m to 10$\times$10 $\mu$m. 
Aperture photometry of the digitized plates was performed with DAOPHOT routines using the same aperture radius and the background annulus as for the CCD photometry. 
The $BVRI$ comparison sequence reported in Semkov et al. (2010) was used as a reference.
Fig. 1 illustrates the quality of the photographic plates used, regardless of their age.

\begin{figure}
   \centering
   \includegraphics[width=8cm]{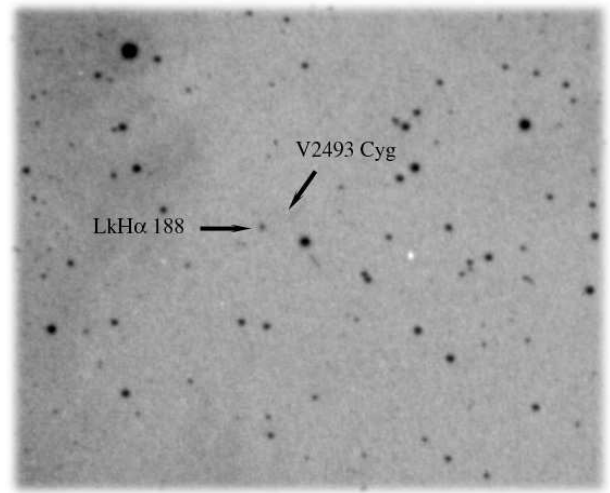}
   \caption{A copy of the photographic plate obtained in the region of NGC 7000 with the first Schmidt telescope in the Hamburg-Bergedorf Observatory on Dec. 18 1932 (exposure: two hours, observer: Bernhard Schmidt). The brightest H$\alpha$ emission star in the field LkH$\alpha$ 188 and the position of V2493 Cyg are marked by arrows. The plate limit is estimated at 17\fm3 (pg). Three of our standard stars (A, B, and C) are visible at the bottom-left of LkH$\alpha$ 188.}
    \end{figure}

The plates from Asiago Schmidt telescope were inspected visually using a high quality Carl Zeiss microscope, which offered a variety of magnifications (Munari et al.  2001). 
The magnitude was then derived by comparing the variable with the stars in the same photometric sequence adopted for the reduction of the CCD observations. 
When ``a" and ``b" are two stars in the sequence and $V$ is the variable, we can estimate the brightness of the variable on photographic plates
inspected visually with a microscope, where ``a" is slightly brighter than $V$ and ``b" slightly fainter. 
The difference in brightness between ``a" and ``b" as perceived by the eye is divided into ten steps, and the difference in magnitude between $V$ and the two stars is counted in terms of these steps, as n1 and n2, conventionally expressed as ``a n1 $V$ n2 b", with n1+n2=10. The magnitude
of the variable $V$ is then obtained directly from the simple arithmetic proportion
         
         mag($V$) = \{[mag(b) - mag(a)]/10\}*n1 + mag(a).

If more than one such pair were available, more estimates would be derived and weighted for the ``a–b", ``c–d" etc. mag interval. 
Typical estimated errors are on the order of 0.10 mag.

The results of the measured magnitudes of V2493 Cyg from the archival photographic plates are given in Table 3. 
The columns provide the name of the observatory, the plate number, the date and JD of observation, photographic emulsions and filters used, the magnitude estimated or plate limit, and the corresponding errors. 
Moreover, V2493 Cyg is invisible (pg$<$17$^{m}$) in the photographic plate obtained by Max Wolf on 1907 Aug. 07 with the 72-cm Waltz-Reflector in the K\"{o}nigsstuhl Observatory, Heidelberg (Germany) (Wenzel 2011).

\onllongtab{3}{
\begin{longtable}{llllllll}
\caption{Data from archival photographic observations of V2493 Cyg during the period 1932-1998.}\\
\hline
\hline
\noalign{\smallskip}
Observatory	&	Plate No.	&	Date	&	J.D.	&	Emulsion	&	Filter	&	Magnitude	&	Er.	\\
\noalign{\smallskip}
\hline
\endfirsthead
\caption{Continued.} \\
\hline
\hline
\noalign{\smallskip}
Observatory	&	Plate No.	&	Date	&	J.D.	&	Emulsion	&	Filter	&	Magnitude	&	Er.	\\
\noalign{\smallskip}
\hline
\endhead
\hline
\endfoot
\hline
\noalign{\smallskip}
\endlastfoot
Hamburg-Bergedorf	&	SS00005	&	1932 Apr 27	&	2426825.293	&	Agfa Isochrom	&	none	&	pg$>$16.5	&	$\pm$0.20	\\
Hamburg-Bergedorf	&	SS00098	&	1932 Sep 25	&	2426976.292	&	Agfa Isochrom	&	none	&	pg$>$17.2	&	$\pm$0.20	\\
Hamburg-Bergedorf	&	SS00116	&	1932 Dec 18	&	2427060.292	&	Agfa Isochrom	&	none	&	pg$>$17.3	&	$\pm$0.20	\\
Palomar	&	001133O	&	1954 Jul 05	&	2434928.874	&	103aO	  &	none	&	pg=19.42	&	$\pm$0.15	\\
Palomar	&	001133E	&	1954 Jul 05	&	2434928.894	&	103aE 	&	Plexi	&	R=16.47	  &	$\pm$0.10	\\
Asiago	&	3028 SP	&	1962 Aug 28	&	2437905.468	&	PanRoy	&	none	&	V$>$14.97	&	$\pm$0.10	\\
Asiago	&	3037 SP	&	1962 Aug 30	&	2437907.446	&	PanRoy	&	none	&	V$>$17.25	&	$\pm$0.10	\\
Asiago	&	3055 SP	&	1962 Sep 06	&	2437913.510	&	PanRoy	&	none	&	V$>$16.04	&	$\pm$0.10	\\
Asiago	&	3101 SP	&	1962 Sep 22	&	2437930.377	&	PanRoy	&	none	&	V$>$17.25	&	$\pm$0.10	\\
Asiago	&	3133 SP	&	1962 Sep 29	&	2437936.529	&	PanRoy	&	none	&	V$>$14.97	&	$\pm$0.10	\\
Asiago	&	3146 SP	&	1962 Sep 30	&	2437938.426	&	PanRoy	&	none	&	V$>$17.25	&	$\pm$0.10	\\
Asiago	&	3161 SP	&	1962 Oct 01	&	2437939.404	&	PanRoy	&	none	&	V$>$17.25	&	$\pm$0.10	\\
Asiago	&	3180 SP	&	1962 Oct 03	&	2437941.332	&	PanRoy	&	none	&	V$>$17.25	&	$\pm$0.10	\\
Asiago	&	3208 SP	&	1962 Oct 17	&	2437955.283	&	PanRoy	&	none	&	V$>$16.04	&	$\pm$0.10	\\
Asiago	&	3219 SP	&	1962 Oct 19	&	2437957.288	&	PanRoy	&	none	&	V$>$17.55	&	$\pm$0.10	\\
Asiago	&	3240 SP	&	1962 Oct 23	&	2437961.279	&	PanRoy	&	none	&	V$>$17.25	&	$\pm$0.10	\\
Asiago	&	3249 SP	&	1962 Oct 24	&	2437962.308	&	PanRoy	&	none	&	V$>$16.04	&	$\pm$0.10	\\
Asiago	&	3294 SP	&	1962 Nov 16	&	2437985.233	&	103a E	&	RG 1	&	R=16.11	  &	$\pm$0.10	\\
Asiago	&	3298 SP	&	1962 Nov 22	&	2437991.249	&	PanRoy	&	none	&	V$>$16.04	&	$\pm$0.10	\\
Asiago	&	3337 SP	&	1962 Nov 26	&	2437995.219	&	PanRoy	&	none	&	V$>$16.04	&	$\pm$0.10	\\
Asiago	&	3347 SP	&	1962 Nov 29	&	2437998.213	&	PanRoy	&	none	&	V$>$17.25	&	$\pm$0.10	\\
Asiago	&	3382 SP	&	1962 Dec 01	&	2438000.236	&	PanRoy	&	none	&	V$>$16.04	&	$\pm$0.10	\\
Asiago	&	3440 SP	&	1962 Dec 14	&	2438013.255	&	PanRoy	&	none	&	V$>$16.04	&	$\pm$0.10	\\
Asiago	&	3483 SP	&	1962 Dec 20	&	2438019.235	&	PanRoy	&	none	&	V$>$16.04	&	$\pm$0.10	\\
Asiago	&	3507 SP	&	1962 Dec 22	&	2438021.212	&	PanRoy	&	none	&	V$>$16.04	&	$\pm$0.10	\\
Asiago	&	4082 SP	&	1963 Jul 18	&	2438229.499	&	PanRoy	&	none	&	V$>$14.97	&	$\pm$0.10	\\
Asiago	&	4095 SP	&	1963 Oct 08	&	2438311.302	&	PanRoy	&	none	&	V$>$17.25	&	$\pm$0.10	\\
Asiago	&	4150 SP	&	1963 Oct 18	&	2438321.308	&	PanRoy	&	none	&	V$>$17.25	&	$\pm$0.10	\\
Asiago	&	4229 SP	&	1963 Nov 08	&	2438342.265	&	PanRoy	&	none	&	V$>$17.25	&	$\pm$0.10	\\
Asiago	&	4238 SP	&	1963 Nov 17	&	2438351.244	&	PanRoy	&	none	&	V$>$17.25	&	$\pm$0.10	\\
Asiago	&	768 SG	&	1967 Sep 17	&	2439750.539	&	1O3aO	  &	none	&	pg$>$17.26	&	$\pm$0.10	\\
Hoher List &	002259	&	1970 May 05	&	2440711.583	&	103aO	&	none	&	pg$>$16.50	&	$\pm$0.20	\\
Hoher List &	002262	&	1970 May 07	&	2440713.510	&	103aO	&	none	&	pg$>$17.20	&	$\pm$0.20	\\
Hoher List &	002270	&	1970 May 08	&	2440714.559	&	103aO	&	none	&	pg$>$17.20	&	$\pm$0.20	\\
Hoher List &	002278	&	1970 May 14	&	2440720.562	&	103aO	&	none	&	pg$>$16.40	&	$\pm$0.20	\\
Hoher List &	002285	&	1970 Jun 01	&	2440739.477	&	103aO	&	none	&	pg$>$17.20	&	$\pm$0.20	\\
Hoher List &	002288	&	1970 Jun 02	&	2440739.545	&	103aO	&	none	&	pg$>$16.50	&	$\pm$0.20	\\
Hoher List &	002289	&	1970 Jun 02	&	2440740.426	&	103aO	&	none	&	pg$>$17.20	&	$\pm$0.20	\\
Hoher List &	002292	&	1970 Jun 02	&	2440740.498	&	103aO	&	none	&	pg$>$17.20	&	$\pm$0.20	\\
Hoher List &	002296	&	1970 Jun 03	&	2440741.437	&	103aO	&	none	&	pg$>$17.20	&	$\pm$0.20	\\
Hoher List &	002299	&	1970 Jun 04	&	2440741.503	&	103aO	&	none	&	pg$>$17.20	&	$\pm$0.20	\\
Asiago	&	4262 SG	&	1971 Mar 25	&	2441035.626	&	103aO	&	GG 13	&	B$>$18.64	&	$\pm$0.10	\\
Asiago	&	8718 SG	&	1971 Mar 25	&	2441035.636	&	103aE	&	RG  1	&	R$>$14.78	&	$\pm$0.20	\\
Asiago	&	4528 SG	&	1971 Jul 31	&	2441163.540	&	103aO	&	GG 13	&	B$>$18.64	&	$\pm$0.10	\\
Asiago	&	9022 SP	&	1971 Sep 18	&	2441213.407	&	TRI X	  &	GG 14	&	V$>$17.23	&	$\pm$0.10	\\
Asiago	&	9122 SP	&	1971 Oct 18	&	2441243.453	&	TRI X	  &	GG 14	&	V$>$16.04	&	$\pm$0.10	\\
Asiago	&	9312 SP	&	1971 Dec 17	&	2441303.297	&	TRI X	  &	GG 14	&	V$>$16.04	&	$\pm$0.10	\\
Byurakan	&	004694	&	1973 Aug 28	&	2441923.280	&	103aO	&	GG13	&	B$>$18.80	&	$\pm$0.10	\\
Byurakan	&	004722	&	1973 Sep 01	&	2441927.291	&	IIaD	  &	GG11	&	V=18.27	&	$\pm$0.14	\\
Byurakan	&	004758	&	1973 Sep 20	&	2441946.348	&	103aO	&	GG13	&	B$>$19.50	&	$\pm$0.10	\\
Byurakan	&	004763	&	1973 Sep 21	&	2441947.229	&	IIaD 	&	GG11	&	V=18.13	&	$\pm$0.14	\\
Asiago	&	6709 SG	&	1973 Oct 18	&	2441974.300	&	103a D	&	GG 14	&	V=17.75	&	$\pm$0.05	\\
Byurakan	&770622A	&	1977 Jun 22	&	2443317.436	&	IIaF	&	RG610	&	R=16.52	&	$\pm$0.10	\\
Byurakan	&	770622B	&	1977 Jun 22	&	2443317.493	&	IIaF	&	RG610	&	R=16.69	&	$\pm$0.10	\\
Asiago	&	10133 SG	&	1979 Aug15	&	2444101.432	&	103a D	&	GG 14	&	V$>$17.25	&	$\pm$0.10	\\
Rozhen (2m)	&	000025	&	1980 Jun 18	&	2444409.485	&	ZU21	&	GG385	&	B$>$18.80	&	$\pm$0.10	\\
Asiago	&	14349 SP	&	1980 Sep 07	&	2444490.477	&	103aO	&	GG 13	&	B$>$16.30	&	$\pm$0.20	\\
Asiago	&	11050 SG	&	1981 Jul 28	&	2444814.502	&	103aD	&	GG 14	&	V=17.60	&	$\pm$0.10	\\
Asiago	&	11084 SG	&	1981 Aug 24	&	2444841.500	&	103aO	&	GG 13	&	B$>$18.64	&	$\pm$0.10	\\
Rozhen (2m)	&	000245	&	1981 Sep 21	&	2444869.274	&	ZU21	&	none	&	pg=19.48	&	$\pm$0.25	\\
Asiago	&	14906 SP	&	1982 Aug 17	&	2445199.488	&	TRI X	&	none	&	V$>$16.04	&	$\pm$0.10	\\
Asiago	&	11682 SG	&	1982 Sep 18	&	2445231.391	&	103aD	&	GG 14	&	V$>$17.25	&	$\pm$0.10	\\
Asiago	&	14982 SP	&	1982 Oct 15	&	2445258.389	&	Tri X	&	none	&	V$>$16.04	&	$\pm$0.10	\\
Asiago	&	14994 SP	&	1982 Oct 16	&	2445259.397	&	Tri X	&	none	&	V$>$16.04	&	$\pm$0.10	\\
Palomar	&	000614  	&	1983 Sep 04	&	2445581.771	&	IIaD	&	W12	&	V=17.66	&	$\pm$0.12	\\
Asiago	&	15354 SP	&	1983 Sep 04	&	2445582.389	&	103aO	&	GG 13	&	B$>$17.02	&	$\pm$0.10	\\
Asiago	&	15361 SP	&	1983 Sep 05	&	2445583.380	&	103aO	&	GG 13	&	B$>$17.27	&	$\pm$0.20	\\
Rozhen (2m)& 000678 &	1983 Sep 12	&	2445590.308	&	ZU21	&	GG385	&	B=19.1	&	$\pm$0.25	\\
Asiago	&	15380 SP	&	1983 Sep 12	&	2445590.411	&	103aO	&	GG 13	&	B$>$17.26	&	$\pm$0.10	\\
Asiago	&	15480 SP	&	1983 Nov 02	&	2445641.389	&	103aO	&	GG 13	&	B$>$17.26	&	$\pm$0.10	\\
Asiago	&	12510 SG	&	1984 Jun 25	&	2445876.588	&	103aO	&	GG 13	&	B$>$16.30	&	$\pm$0.20	\\
Asiago	&	12519 SG	&	1984 Jul 04	&	2445886.498	&	103aO	&	GG 13	&	B$>$18.64	&	$\pm$0.10	\\
Asiago	&	16379 SP	&	1985 Aug 13	&	2446290.503	&	103aO	&	GG 13	&	B$>$17.26	&	$\pm$0.10	\\
Asiago	&	14580 SG	&	1989 Aug 02	&	2447740.534	&	IN	  &	RG 5	&	I=14.73	  &	$\pm$0.10	\\
Asiago	&	14638 SG	&	1989 Oct 06	&	2447806.458	&	103aO	&	GG 13	&	B$>$17.26	&	$\pm$0.10	\\
Asiago	&	14639 SG	&	1989 Oct 06	&	2447806.488	&	IN	&	RG 8	&	I=14.73	&	$\pm$0.10	\\
Asiago	&	14657 SG	&	1989 Oct 25	&	2447825.338	&	103aO	&	GG 13	&	B$>$18.64	&	$\pm$0.10	\\
Asiago	&	14658 SG	&	1989 Oct 25	&	2447825.370	&	IN	&	RG 8	&	I=15.11	&	$\pm$0.10	\\
Asiago	&	14719 SG	&	1989 Nov 29	&	2447860.330	&	103aO	&	GG 13	&	B$>$18.64	&	$\pm$0.10	\\
Asiago	&	14720 SG	&	1989 Nov 29	&	2447860.358	&	IN	&	RG 8	&	I=14.54	&	$\pm$0.10	\\
Asiago	&	14810 SG	&	1990 Jul 27	&	2448100.465	&	103aO	&	GG 13	&	B$>$18.95	&	$\pm$0.10	\\
Asiago	&	14811 SG	&	1990 Jul 27	&	2448100.496	&	IN	&	RG 8	&	I=14.92	&	$\pm$0.10	\\
Asiago	&	14859 SG	&	1990 Aug 23	&	2448127.494	&	103aO	&	GG 13	&	B$>$18.95	&	$\pm$0.10	\\
Asiago	&	14870 SG	&	1990 Aug 26	&	2448129.526	&	INsen	&	RG 8	&	I=15.01	&	$\pm$0.10	\\
Palomar	&	003536  	&	1990 Sep 11	&	2448145.728	&	IIIaF	&	RG610	&	R=16.29	&	$\pm$0.10	\\
Asiago	&	14904 SG	&	1990 Oct 22	&	2448187.368	&	INsen	&	RG 8	&	I=14.82	&	$\pm$0.10	\\
Rozhen (Sch)&	005738&	1991 Jan 11	&	2448266.610	&	ZU21	&	GG385	&	B$>$18.50	&	$\pm$0.20	\\
Asiago	&	18303 SP	&	1991 Oct 09	&	2448539.469	&	103aO	&	GG 13	&	B$>$16.47	&	$\pm$0.10	\\
Asiago	&	15062 SG	&	1991 Nov 07	&	2448568.386	&	103aO	&	GG 13	&	B$>$17.26	&	$\pm$0.10	\\
Asiago	&	15063 SG	&	1991 Nov 07	&	2448568.414	&	INsen	&	RG 8	&	I$>$14.63	&	$\pm$0.10	\\
Asiago	&	15072 SG	&	1991 Nov 30	&	2448591.411	&	103aO	&	GG 13	&	B$>$18.64	&	$\pm$0.10	\\
Asiago	&	15073 SG	&	1991 Nov 30	&	2448591.447	&	INsen	&	RG 8	&	I=14.92	  &	$\pm$0.20	\\
Asiago	&	15083 SG	&	1991 Dec 02	&	2448593.416	&	103aO	&	GG 13	&	B$>$17.26	&	$\pm$0.10	\\
Asiago	&	15084 SG	&	1991 Dec 02	&	2448593.446	&	INsen	&	RG 8	&	I$>$14.63	&	$\pm$0.10	\\
Asiago	&	15106 SG	&	1991 Dec 06	&	2448597.313	&	103aO	&	GG 13	&	B$>$18.95	&	$\pm$0.10	\\
Asiago	&	15107 SG	&	1991 Dec 06	&	2448597.341	&	INsen	&	RG 8	&	I=14.73	  &	$\pm$0.10	\\
Asiago	&	15252 SG	&	1992 Jul 02	&	2448806.496	&	INsen	&	RG 8	&	I=14.55	  &	$\pm$0.10	\\
Asiago	&	15271 SG	&	1992 Aug 27	&	2448862.458	&	103aO	&	GG 13	&	B$>$18.95	&	$\pm$0.10	\\
Asiago	&	15290 SG	&	1992 Sep 25	&	2448891.475	&	103aO	&	GG 13	&	B$>$18.95	&	$\pm$0.10	\\
Asiago	&	15291 SG	&	1992 Sep 26	&	2448891.504	&	INsen	&	RG 8	&	I=14.73	  &	$\pm$0.10	\\
Asiago	&	15304 SG	&	1992 Oct 24	&	2448920.372	&	103aO	&	GG 13	&	B$>$18.95	&	$\pm$0.10	\\
Asiago	&	15305 SG	&	1992 Oct 24	&	2448920.404	&	INsen	&	RG 8	&	I=15.68	  &	$\pm$0.20	\\
Asiago	&	15341 SG	&	1992 Dec 18	&	2448975.291	&	103aO	&	GG 13	&	B$>$18.64	&	$\pm$0.20	\\
Asiago	&	15342 SG	&	1992 Dec 18	&	2448975.322	&	INsen	&	RG 8	&	I=15.68	  &	$\pm$0.20	\\
Palomar	&	005254  	&	1993 Jun 27	&	2449165.910	&	IIIaJ	&	GG385	&	B=19.09	  &	$\pm$0.15	\\
Asiago	&	15641 SG	&	1993 Sep 10	&	2449241.415	&	103aO	&	GG 13	&	B$>$18.95	&	$\pm$0.10	\\
Asiago	&	15642 SG	&	1993 Sep 10	&	2449241.449	&	IN  	&	RG 8	&	I=14.61	  &	$\pm$0.10	\\
Asiago	&	15657 SG	&	1993 Oct 19	&	2449280.414	&	103aO	&	GG 13	&	B$>$18.64	&	$\pm$0.10	\\
Asiago	&	15671 SG	&	1993 Nov 18	&	2449310.373	&	103aO	&	GG 13	&	B$>$18.95	&	$\pm$0.10	\\
Asiago	&	15672 SG	&	1993 Nov 18	&	2449310.401	&	IN	  &	RG 8	&	I=14.56	  &	$\pm$0.10	\\
Asiago	&	15684 SG	&	1993 Dec 09	&	2449331.349	&	103aO	&	GG 13	&	B$>$18.64	&	$\pm$0.20	\\
Asiago	&	15685 SG	&	1993 Dec 09	&	2449331.384	&	IN	  &	RG 8	&	I=14.73  	&	$\pm$0.10	\\
Asiago	&	15690 SG	&	1993 Dec 10	&	2449332.257	&	103aO	&	GG 13	&	B$>$18.95	&	$\pm$0.10	\\
Asiago	&	15786 SG	&	1994 Aug 12	&	2449577.379	&	103aO	&	GG 13	&	B$>$18.64	&	$\pm$0.10	\\
Asiago	&	15787 SG	&	1994 Aug 12	&	2449577.407	&	IR H	&	RG 8	&	I$>$13.30	&	$\pm$0.10	\\
Palomar	&	005993	  &	1994 Sep 12	&	2449607.686	&	IVN	  &	RG9	  &	I=15.12	  &	$\pm$0.10	\\
Asiago	&	15821 SG	&	1994 Nov 28	&	2449685.378	&	103aO	&	GG 13	&	B$>$18.64	&	$\pm$0.10	\\
Asiago	&	16003 SG	&	1995 Nov 22	&	2450044.350	&	103aO	&	GG 13	&	B$>$18.64	&	$\pm$0.10	\\
Asiago	&	16004 SG	&	1995 Nov 22	&	2450044.378	&	IR H	&	RG 8	&	I=15.11	  &	$\pm$0.10	\\
Asiago	&	16124 SG	&	1996 Jul 11	&	2450275.532	&	103aO	&	GG 13	&	B$>$18.64	&	$\pm$0.10	\\
Asiago	&	16125 SG	&	1996 Jul 11	&	2450275.576	&	IN	  &	RG 8	&	I$>$15.58	&	$\pm$0.10	\\
Asiago	&	16147 SG	&	1996 Sep 08	&	2450334.526	&	103aO	&	GG 13	&	B$>$18.95	&	$\pm$0.10	\\
Asiago	&	16148 SG	&	1996 Sep 08	&	2450334.583	&	IN	  &	RG 8	&	I=14.62	  &	$\pm$0.10	\\
Asiago	&	16158 SG	&	1996 Oct 04	&	2450361.348	&	IN	  &	RG 8	&	I=14.62	  &	$\pm$0.10	\\
Asiago	&	16166 SG	&	1996 Nov 04	&	2450392.283	&	103aO	&	GG 13	&	B$>$18.64	&	$\pm$0.10	\\
Asiago	&	16167 SG	&	1996 Nov 04	&	2450392.331	&	IN  	&	RG 8	&	I=15.11	  &	$\pm$0.10	\\
Asiago	&	16175 SG	&	1996 Nov 08	&	2450396.314	&	103aO	&	GG 13	&	B$>$18.95	&	$\pm$0.10	\\
Asiago	&	16176 SG	&	1996 Nov 08	&	2450396.341	&	IN  	&	RG 8	&	I=14.45	  &	$\pm$0.05	\\
Asiago	&	16182 SG	&	1996 Dec 04	&	2450422.263	&	103aO	&	GG 13	&	B$>$18.95	&	$\pm$0.10	\\
Asiago	&	16467 SG	&	1997 Aug 27	&	2450688.498	&	103aO	&	GG 13	&	B$>$18.64	&	$\pm$0.10	\\
Asiago	&	16468 SG	&	1997 Aug 28	&	2450688.539	&	INsen	&	RG 8	&	I=15.11  	&	$\pm$0.10	\\
Asiago	&	16507 SG	&	1997 Oct 01	&	2450723.412	&	103aO	&	GG 13	&	B$>$18.95	&	$\pm$0.10	\\
Asiago	&	16508 SG	&	1997 Oct 01	&	2450723.441	&	IN  	&	RG 8	&	I=14.62  	&	$\pm$0.10	\\
Asiago	&	16705 SG	&	1998 Oct 22	&	2451109.345	&	TP4415S	&	BG 12	&	B$>$17.26	&	$\pm$0.10	\\
Asiago	&	16720 SG	&	1998 Nov 17	&	2451135.389	&	TP4415S	&	BG 12	&	B$>$17.26	&	$\pm$0.10	\\
Asiago	&	16726 SG	&	1998 Nov 18	&	2451136.399	&	TP4415S	&	BG 12	&	B$>$17.26	&	$\pm$0.10	\\
\end{longtable}
}
   
\section{Results and discussion}

Photometric data presented in this paper show the usefulness of systematically monitoring the star forming regions.
These data can be used to detect new FUor or EXor events and to determine the type of the outburst.
The $BVRI$ light curves of V2493 Cyg are plotted in Fig. 1. 
The filled diamonds represent the CCD observations from the present paper,
the filled circles observations from the 48 inch Samuel Oschin telescope at Palomar Observatory (Miller et al. 2011), 
the open diamonds photographic data from the Asiago Schmidt telescopes,
the open squares photographic data from the Palomar Schmidt telescope,
the filled squares photographic data from the Byurakan Schmidt telescope
and the open circles photographic data from the Rozhen 2-m RCC telescope.
The optical photometric data published by other authors (K{\'o}sp{\'a}l et al. 2011; Lorenzetti et al. 2012) show the same changes in brightness during the outburst, which are perfectly matched with the light curves shown in Fig 1.

\begin{figure*}
   \centering
   \includegraphics[width=17cm]{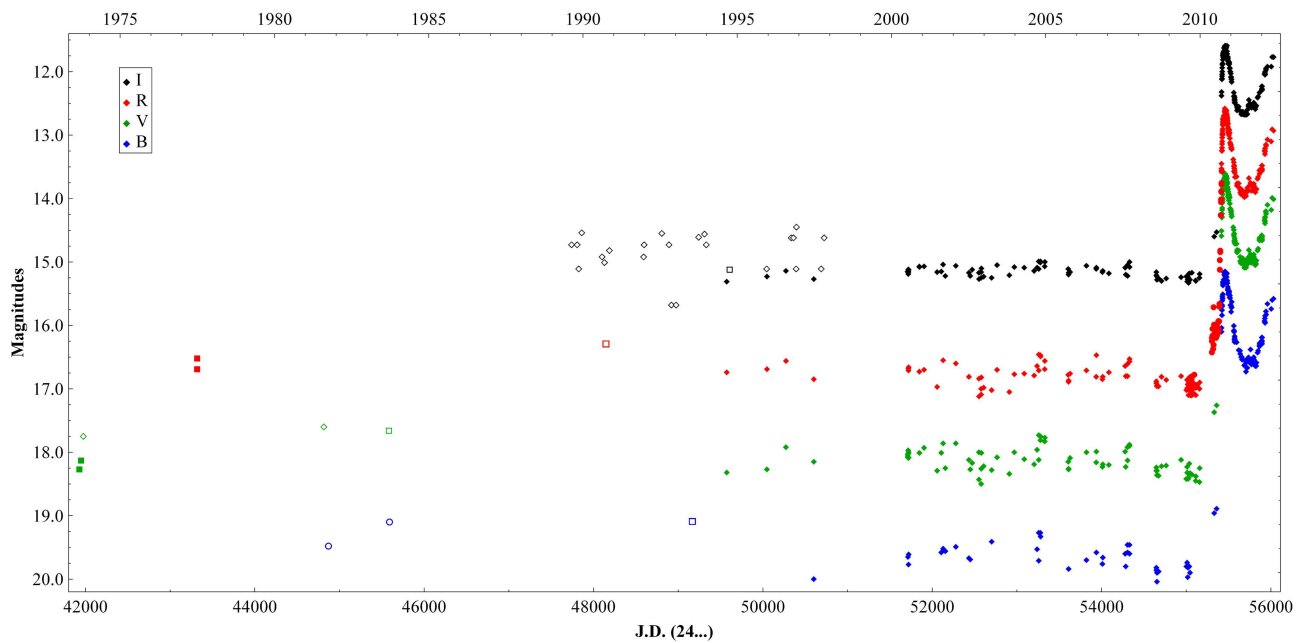}
      \caption{Historical $BVRI$ light curves of V2493 Cyg for the period September 1973 $-$ April 2012.}
         \label{Fig1}
  \end{figure*}

The historical light curves of V 2493 Cyg allows us to conclude that its photometric behavior is similar to that of a classical T Tauri star. 
The same conclusion was made by Lorenzetti et al. (2011) based on a comparison of infrared photometry from Cohen \& Kuhi (1979) and from the 2MASS catalog.
Other large-amplitude eruptions have not yet been registered in our long-term photometric study, but due to the sparse and random photometry available at this stage, the occurrence of these short events cannot be ruled out.
After reaching its maximum brightness in October 2010 (Semkov et al. 2010), the brightness of V2493 Cyg declined slowly, having weakened by 1\fm45 ($V$) by the spring/early summer of 2011.
During the summer of 2011, V2493 Cyg underwent only minor brightness variations around the mean level that was brighter than the quiescence level by 3\fm3 ($V$). 
From October 2011, another light increase occurred and the star became brighter by 1\fm1 ($V$) until April 2012. 
Consequently, the outburst of V2493 Cyg continued for approximately two years. 

Fig. 3 represents the change in position of V2493 Cyg during the outburst on the color-color diagram $V-R/R-I$.
The location of main-sequence dwarfs (blue line) according to Johnson (1966) and the interstellar reddening vector for the region A$_{V}$ = 3\fm4 (black line) according to Cohen \& Kuhi (1979) are shown. 
Data for the main-sequence stars are recalculated to correspond to the Johnson's to Cousins system using the corresponding equations in Moro \& Munari (2000).
Fig. 3 shows that V2493 Cyg becomes appreciably bluer when its brightness increases during the outburst.

\begin{figure}
   \centering
   \includegraphics[width=8cm]{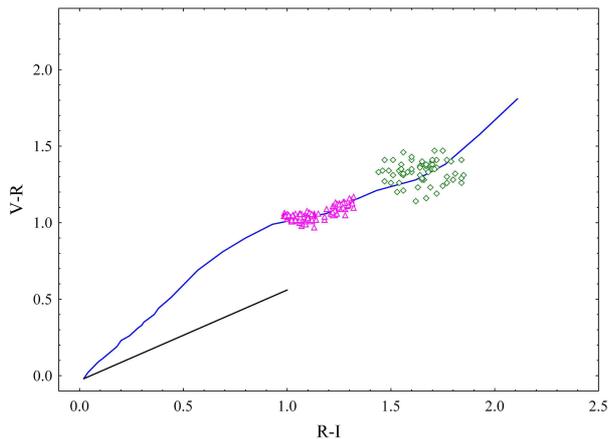}
   \caption{$V-R$ vs. $R-I$ color-color diagram for V2493 Cyg before (open diamonds) and during the outburst (open triangles). The solid blue line
is the locus of main-sequence dwarf stars. The straight black line is the reddening vector for $A_{V}$ = 3\fm4.}
    \end{figure}

The spectral observations obtained during the past few months show substantial changes in the profile of H$\alpha$ line.
Fig. 4 presents a comparison between the low-resolution spectra of V2493 Cyg obtained at the beginning of outburst (September 2010) and during the recent increase in brightness (December 2011).
In the first spectrum, the H$\alpha$ line is seen in emission, which is confirmed by the low-resolution spectroscopy of Miller et al. (2011).
The second spectrum already clearly shows the strong P Cyg shape of the H$\alpha$ line. 
A significant change in the SED of V2493 Cyg relative to both spectra is observed, as also confirmed by the photometric data (Table 1).
During the second rise in brightness, the star is significantly bluer than during the first.

\begin{figure}
   \centering
   \includegraphics[width=9cm]{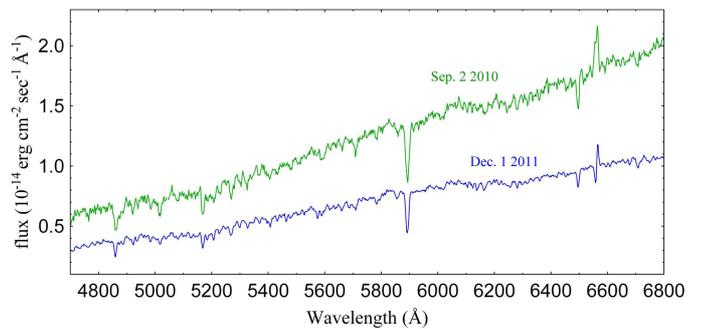}
   \caption{Low-resolution spectra of V2493 Cyg obtained on Sep 2, 2010 (the 60-cm telescope of Schiaparelli Observatory in Varese) and on Dec 1, 2011 (1.22-m telescope of Asiago Observatory).}
    \end{figure}

From the high-resolution spectroscopy, the P Cyg profile of the H$\alpha$ line was detectable at the beginning of the outburst (Miller et al. 2011; Lee et al. 2011).
The deep and high-velocity blueshifted absorption is interpreted as evidence of a strong outflow driven by the FUor object. 
According to Miller et al. (2011), the blueshifted absorption of H$\alpha$ line extends up to $\sim$-200 km/s on September 2010. 
Our most recent high and medium resolution spectroscopy (Table 2) indicate that there has been a significant increase in the velocity of the absorption component of H$\alpha$ and Na I D lines (Fig. 5).
The terminal velocity of the wind during the two months covered by our high resolution monitoring remained stable at about $-$500 km/s, while profound changes were observed in the shape of the absorption component.

\begin{figure}
   \centering
   \includegraphics[width=8cm]{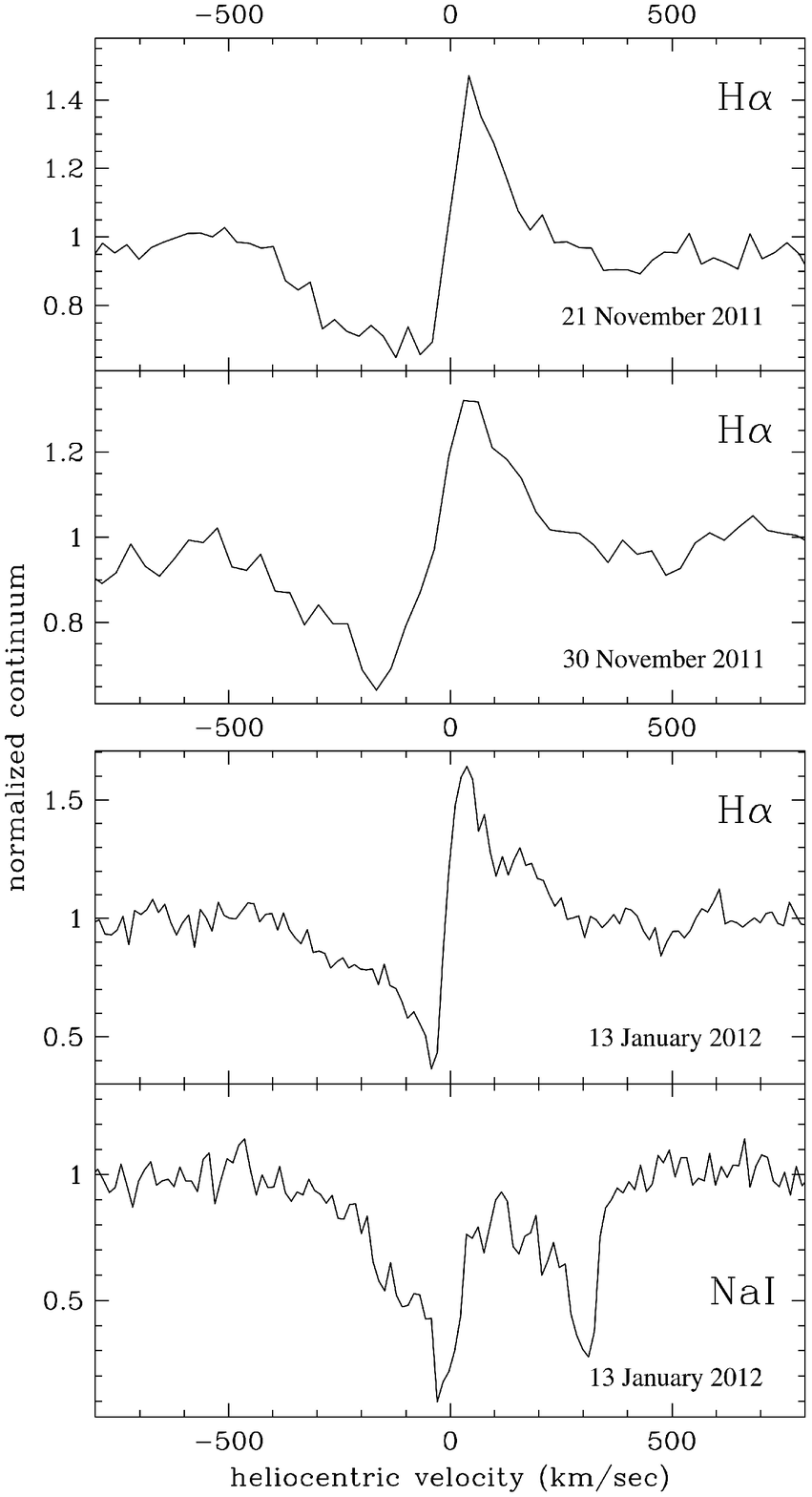}
   \caption{The profiles of H$\alpha$ and Na I D lines extracted from the three spectra of V2493 Cyg obtained during the second increase in brightness (cf Table~2).}
    \end{figure}    

The interpretation of the profile changes of the absorption component is complicated by the uncertainties in the location and the precise mechanism responsible for the formation of the line, whichmay be a spherically symmetric wind, blowing from either the stellar surface or a circumstellar disk, that is radiatively accelerated, etc.  
The changes observed in the profiles of Fig. 5 could be for example accounted for by a variation in the optical depth in the line.  
In this framework, the profile for Nov 21, 2011 would correspond to an optical depth larger than that measured on Jan 13, 2012, with the profile for Nov 30, 2011 being characterized by an in-between value of the optical depth.  
There has been no noticeable correlation between the variations in the optical depth of H$\alpha$ line and the photometric properties during recent months.
Such a correlation is also not found in other classical FUors that appear to have a highly variable stellar wind and no significant changes the brightness (Herbig et al. 2003; Herbig 2009).

To classify the absorption spectrum of V2493 Cyg, we compared our spectra with the Asiago atlas of MKK spectral types observed with exactly the same    instrumental configuration of V2493 Cyg (Munari 2012, in preparation).
We took the absolutely fluxed spectra of V2493 Cyg and the MKK atlas and continuum normalized them using the same function (a Legendre polynomial of fifth order limited to the range of wavelength covered in Fig. 6 which correspond to those recommended for the classification within the MKK system). 
As a first classification pass, we applied a simple $\chi^2$ matching to determine the area of the Herzsprung-Russell diagram on which our deeper analysis was focused. 
The match found by the $\chi^2$ is not perfect, since the stellar spectra originates in stationary atmospheres where a three-dimensional treatment is generally unnecessary, while the absorption lines in V2493 Cyg instead form in a moving medium, the wind. 
We then proceeded to refine the classification by using an eye inspection of the spectra, and found that the closest (even though imperfect) match was for a G3I type star. 
Fig. 6 shows how the properties of the absorption spectrum of V2493 Cyg are in-between those of supergiants of the G0 and G5 types.

\begin{figure*}
   \centering
   \includegraphics[width=17cm]{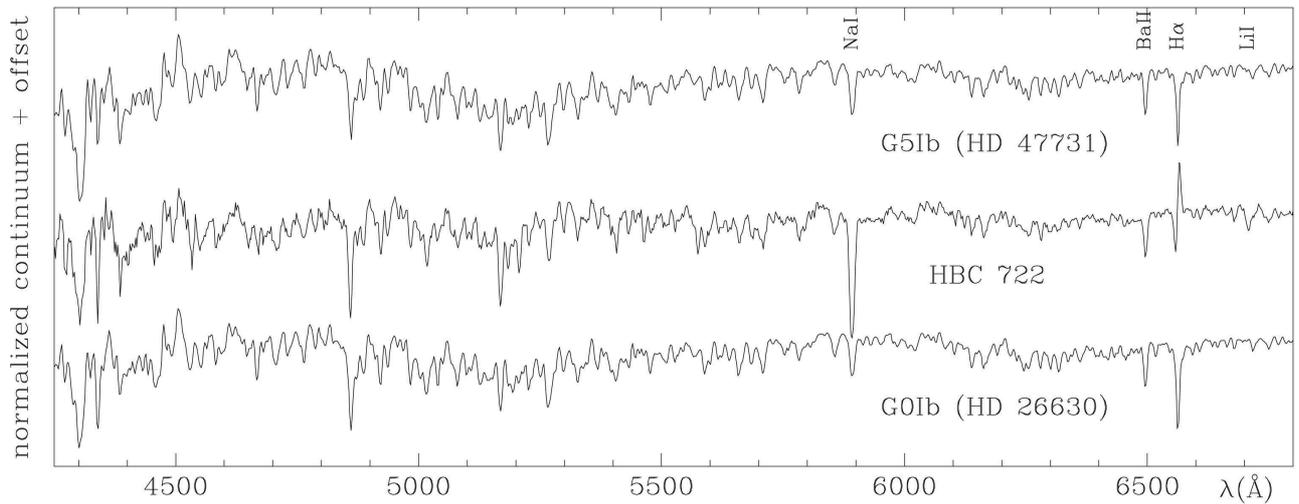}
   \caption{High-resolution spectrum of V2493 Cyg (HBC 722) is compared with spectra of the G supergiant stars HD 47731 and HD 26630 (from Asiago spectral database).}
    \end{figure*}   

All published studies of V2493 Cyg eruption tried to determine whether it is either a FUor or EXor type star. 
Most authors (Semkov et al. 2010; Miller et al. 2011; Green et al. 2011; Lee et al. 2011; Pooley \& Green 2010; Reipurth et al. 2012) classified V2493 Cyg as a typical FUor, but others (K{\'o}sp{\'a}l et al. 2011; Leoni et al. 2010, Lorenzetti et al. 2012) considered the observed outburst to have instead a EXor type.
Both types of PMS eruptive stars have been classified in terms of their wide range of available photometric and spectral properties, but their outbursts are thought to have the same cause - an enhanced accretion rate. 
The observed eruption of V1647 Ori appears to have characteristics similar to those of FUor and EXor, and it is assumed to be in-between the two types. 
According to Aspin (2011), the viewing inclination angle of the star/disk system can play a significant role in the observed spectral features. Therefore, it is assumed that the two types of outbursts may be much closer in nature.

Unlike the classical FUor stars, V2493 Cyg has a relatively low outburst luminosity $\sim$12$L_{\sun}$ (Miller et al. 2011) and low accretion rate $\sim$10$^{-6}$$M_{\sun}$$/$yr (K{\'o}sp{\'a}l et al. 2011). 
Most of the FUor objects have luminosities on the order of 100$L_{\sun}$ during the maximum. 
The luminosity of V2493 Cyg is at the faint end of the luminosity range of FUor outbursts but still comparable to those of some FUor objects such as L1551 IRS5 and HH381 IRS (Reipurth \& Aspin 2010). 

From all objects associated with the group of FUors, only three (FU Ori, V 1057 Cyg, and V 1515 Cyg) have detailed photometric observations taken during the outburst and during the set of brightness (Clarke et al. 2005).
For some years, we have made efforts to construct the historical light curves of several other FUor and FUor-like objects such as: V1735 Cyg (Peneva et al. 2009), Parsamian 21 (Semkov \& Peneva 2010c), V733 Cep (Peneva et al. 2010), and V582 Aur (Semkov et al. 2011).
Our results suggest that each object of this type has a characteristic long-term light curve, which distinguishes it from other objects. 
The shape of the observed light curves of FUors may vary considerably in the time of rise, the rate of decrease in brightness, the time spent at maximum brightness, and the light variability during the set in brightness. 

The light curve of V2493 Cyg from all available photometric observations is also somewhat unique.
The rate of increase in the brightness (the fastest ever recorded) was followed by a very rapid fall in brightness.
But the most remarkable feature of the light curve of V2493 Cyg is the repeated rise in brightness in the past few months. 
At present, it is impossible to predict how long the rise in brightness will continue and which maximal stellar magnitude will be exceeded.

\section{Concusions}

On the basis of our photometric monitoring over the past two years and the spectral properties at the maximal light (a G3I supergiant spectrum with strong P Cyg profiles of H$\alpha$ and Na I D lines), we have confirmed that the observed outburst of V2493 Cyg is of FUor type.
V2493 Cyg was the first FUor object observed in all spectral ranges during the rise of the brightness as well as during the first year after reaching the maximum brightness. 
At the same time, according to existing observations the light curve of V2493 Cyg remains unique, confirming the hypothesis that each known FUor has a different rate of increase and decrease in brightness, and different light curve shape.
We plan to continue our spectroscopic and photometric monitoring of the star during the next few months and years and strongly encourage similar follow-up observations.

\begin{acknowledgements}
     This work was partly supported by grants DO 02-85, DO 02-273, DO-02-275, and DO 02-362 of the National Science Fund of the Ministry of Education, Youth and Science, Bulgaria. 
     The authors thank the Director of Skinakas Observatory Prof. I. Papamastorakis and Prof. I. Papadakis for the award of telescope time and to Ivana Poljan\v{c}i\'{c} for the assistance in measuring Asiago plates. 
     We also thanks Derlef Groote for providing access to the Hamburg observatory digitized plate collection. 
     The Digitized Sky Survey was produced at the Space Telescope Science Institute under U.S. Government grant NAG W-2166. 
     The images of these surveys are based on photographic data obtained using the Oschin Schmidt Telescope on Palomar Mountain and the UK Schmidt Telescope. 
     The plates were processed into the present compressed digital form with the permission of these institutions. 
     This research has made use of the NASA Astrophysics Data System.
\end{acknowledgements}

\end{document}